# The Robust Digital Image Watermarking using Quantization and Fuzzy Logic Approach in DWT Domain

[1]Nallagarla Ramamurthy, [2]Dr.S.Varadarajan

[1] Research Scholar, JNTUA, Anantapur, INDIA

[2] Professor, Dept. of ECE, S.V. University College of Eng. , Tirupati, INDIA

### Abstract

In this paper a novel approach to embed watermark into the host image using quantization with the help of Dynamic Fuzzy Inference System (DFIS) is proposed. The cover image is decomposed up to 3- levels using quantization and Discrete Wavelet Transform (DWT). A bitmap of size 64x64 pixels is embedded into the host image using DFIS rule base. The DFIS is utilized to generate the watermark weighting function to embed the imperceptible watermark. The implemented watermarking algorithm is imperceptible and robust to some normal attacks such as JPEG Compression, salt&pepper noise, median filtering, rotation and cropping.

## 1. Introduction

The transmission of multimedia data became daily routine nowadays and it is necessary to find an efficient way to transmit through various networks. Copyright protection of multimedia data has become critical issue due to massive spreading of broadband networks, easy copying, and new developments in digital technology [1]. As a solution to this problem, digital image watermarking became very popular nowadays. Digital image watermarking is a kind of technology that embeds copyright information into multimedia content. An effective image watermarking mainly includes watermark generation, watermark embedding, watermark detection, and watermark attack [5], [1]. Digital image watermarking provides copyright protection to image by hiding appropriate information in original image to declare rightful ownership [6]. There are four essential factors those are commonly used to determine quality of watermarking scheme. They are robustness, imperceptibility, capacity, and blindness. Robustness is a measure of immunity of watermark against attempts to image modification and manipulation like compression, filtering, rotation, scaling, noise attacks, resizing, cropping etc. Imperceptibility is the quality that the cover image should   not be destroyed by the presence of watermark.  Capacity includes techniques that make it possible to embed majority of information. Extraction of watermark from watermarked image without the need of original image is referred to as blind watermarking. The non-blind watermarking technique requires that the original image to exist for detection and extraction. The semi-blind watermarking scheme requires the secrete key and watermark bit sequence for extraction. Another categorization of watermarks based on the embedded data is visible or invisible [7].

According to the domain of watermark insertion, the watermarking techniques fall into two categories:  spatial domain methods and transform domain methods. Many techniques have been proposed in the spatial domain such as LSB (Least Significant Bit) insertion method, the patch work method and the texture block coding method [8]. These techniques process the location and luminance of the image pixel directly. The LSB method has a major disadvantage that the least significant bits may be easily destroyed by lossy compression. Transform domain method based on special transformations, and process the coefficients in frequency domain to hide the data. Transform domain methods include Fast Fourier Transform(FFT), Discrete Cosine Transform(DCT), Discrete Wavelet Transform(DWT), Curvelet Transform(CT), Counterlet Transform(CLT) etc. In these methods the watermark is hidden in the high and middle  frequency coefficients of the cover image. The low frequency coefficients are suppressed by filtering as noise, hence watermark is not inserted in low frequency coefficients [8]. The transform domain method is more robust than the spatial domain method against compression, filtering, rotation, cropping and noise attack etc.

In [9], Yanhong Zhang proposed a blind watermarking algorithm using the RBF neural network. In [1], we proposed the robust digital image watermarking using quantization and Back Propagation Neural Network (BPNN), where the experimental results show that our algorithm is better than the algorithm proposed in [9].
In this paper, a novel robust digital image watermarking algorithm is presented to insert the watermark in blue plane of the cover image using quantization and DFIS. The fuzzy model is exploited to determine a valid approximation of the quantization step.

## 2. Discrete Wavelet Transform (DWT)

The Discrete Wavelet Transform (DWT) provides both spatial and frequency description of an image. Unlike conventional Fourier transform, temporal information is retained in this transformation process [10]. Wavelets are created by translations and dilations of a fixed function called mother wavelet.

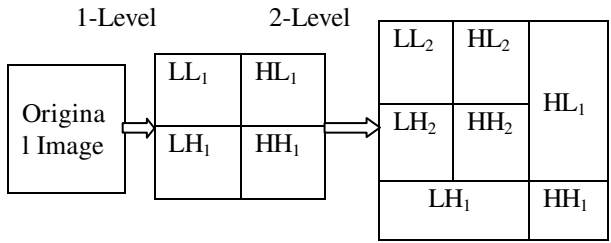

Figure (1): Discrete Wavelet Transformation

The Wavelet transform decomposes the image into three spatial directions: horizontal, vertical and diagonal. The multi-resolution of wavelet allows representing an image at more than one resolution level. The magnitude of DWT coefficients is larger in the lowest bands (LL) and is smaller in other bands HH, LH and HL, at each level of decomposition. High resolution sub bands help to easily locate edge and texture patterns in an image. Wavelet transform can accurately model Human Visual System (HVS), compared to other transforms like DFT and DCT. This allows embedding higher energy watermarks in regions, where HVS is less sensitive. Embedding watermark in these regions allow us to increase robustness of watermark without damaging image fidelity. DWT provides multi-resolution of an image, so that the image can be sequentially processed from low resolution to high resolution. The advantage of this approach is that the features of an image that might not be detected at one resolution can easily be detected at another resolution.

## 3. Dynamic Fuzzy Inference System (DFIS)

The Dynamic Fuzzy Inference System (DFIS), also known

as Dynamic Fuzzy Expert System, is a widely accepted computing framework based on the popular concepts of fuzzy set theory, fuzzy if-then rules and fuzzy reasoning [3], [9]. The DFIS is recognized to provide simple fuzzy approaches in order to perform the mapping from a given set of inputs to another set of outputs without the extensive use of mathematical modeling concepts. In general, a DFIS is composed of four different function blocks namely, a fuzzifier, a knowledge base, a fuzzy inference engine and a defuzzifier as shown in figure(2).

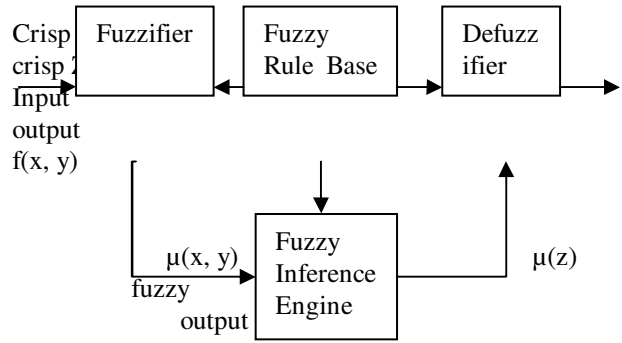

Figure (2): Dynamic Fuzzy Inference System

The Fuzzifier transfer crisp inputs into fuzzy sets. The Knowledge Base encompasses a database and rule base. The data base defines the membership functions of the linguistic variables. The rule base consists of a set of IF-THEN rules that can be given by a human expert or also can be extracted from the linguistic description of the data. A typical if- then rule has a premise as well as conclusion. In this Mamdani type DFIS; the fuzzy rule base has the following form: If input is 'f' then output is 'z', where 'f' is fuzzy input and 'z' is quantized output. The inference Engine is a general control mechanism that exploits the fuzzy rules and the fuzzy sets defined in the Knowledge Base in order to reach certain conclusion. The Defuzzifizer is used to convert fuzzy outputs of the fuzzy rules into crisp output values. The Mamdani type DFIS model is suited to represent the behavior of a non-linear system as it interpolates between multiple linear models. Therefore the Mamdani type DFIS is ideal to model the watermark weighting function, as it incorporates the fuzzy and nonlinear aspect of human vision.

## 4. Watermark Embedding

In this paper, the robust digital image watermarking scheme is developed based on DWT and DFIS. The primary novelty of this scheme is that the Mamdani type DFIS model is exploited in order to determine a valid approximation of a quantization step of each DWT coefficient. Furthermore, the HVS properties are modeled using Haar wavelet to improve watermark robustness and imperceptibility. This information is utilized by the algorithm to generate the watermark weighting function that would enable the robust and imperceptible watermark. The basic concept underlying fuzzy logic is that variable values are words or linguistic variables rather than numbers, their use is closer to human intuition. Computing with words exploits the tolerance for imprecision and there by lowers the cost of solution. Watermarking based on fuzzy logic is developed to extract human eye sensitivity knowledge. Haar wavelet generate two random parts of watermark, one is embedded in cover image and the other is kept as a secrete key for watermark extraction. The watermark is embedded into high and middle frequency sub bands of the wavelet transform, even though this can clearly change the image fidelity. The advantage of HVS model is that the watermark is embedded carefully without degrading image perceptibility. The HVS utilizes the texture sensitivity and multi resolution structure of the wavelets to embed the watermark without degrading the perceptibility of the watermarked image. The invisibility of the watermark in watermarked image is determined by an observer at a distance equal to 6 times the size of the image.

The imperceptibility of watermarked image is proportional to the texture sensitivity of an image. The texture sensitivity can be estimated by quantizing DWT coefficients of an image using quantization values. The result is then rounded to the nearest integer. The number of non-zero coefficients is then calculated. The texture sensitivity can be calculated by the following formula.

$$S_j = \sum_j round\left[(T_{j+key})/Q\right] \quad \text{...... (1)}$$

Where, round [(Tj+key)/Q] takes the rounded value of [(Tj+key)/Q] and returns '1' if the value is not equal to zero, otherwise returns '0'.

The inference system uses a set of rules which are primarily based on the facts:

1. The eye is less sensitive to noise in those areas of the image where brightness is high is high or low.

2. The eye is less sensitive to noise in highly textured areas, but among these, more sensitive near the edges.

3. The eye is less sensitive to noise in the regions with high brightness and changes in very dark regions.

In this work rules are developed based on texture sensitivity as follows

If input is low then output is low
If input is medium then output is medium
If input is high then output is high

The overall output of the system will be obtained as a weighted average of al rule outputs as follows

$$O_i = \frac{\sum_i^n w_i z_i}{\sum_i w_i} \quad \text{...... (2)}$$

Where 'corresponds to the number of rules depicted, 'w_i'is the fringing strength that weights output 'z_i'of rule 'i'.

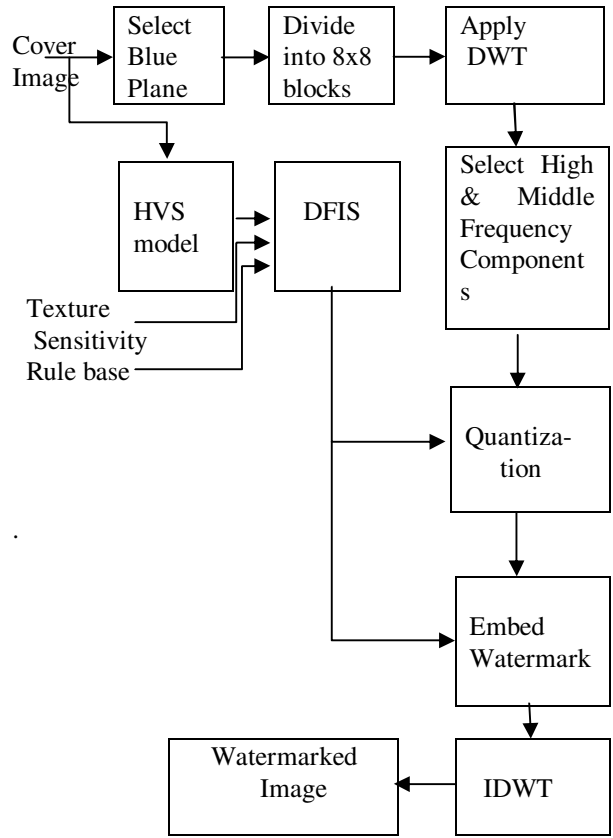

Figure (3): Watermark Embedding

**Watermark Embedding Algorithm:**

1. Read a color image of size NxN.

2. Resize the color image to 512x512 pixels and select Blue plane to embed watermark.

3. Select bitmap of 64x64 as watermark.

4. Perform 3- level DWT on cover image to obtain the frequency sub components {$HH_1$, HL1, LH1, {$HH_2$, HL2, LH2}, HH3, HL3, LH3}}}.

5. Select the beginning position of to embed watermark by generating secrete key.

6. Compute the texture sensitivity of the selected components to embed watermark, and apply these coefficients to DFIS.

7. Apply fuzzy inference rules to DFIS to generate watermark weighting factor.

8.  Perform watermark embedding in low frequency DWT coefficients of the host image using the following formula

$$T'_{j+key} = DFIS(\sum_j round(\frac{T_{j+key}}{Q})) + X_j \quad .... (3)$$

Where
$X_j$ is watermark sequence
Q is quantization value
$T'_{j+key}$ is coefficient of watermarked image

9.  Perform IDWT on each coefficient to get watermarked image.

## 5. Watermark Extraction

The watermark extraction is the reverse process of that of watermark embedding. The watermark is extracted by taking the difference of DWT coefficients of watermarked image and the output coefficients of DFIS.

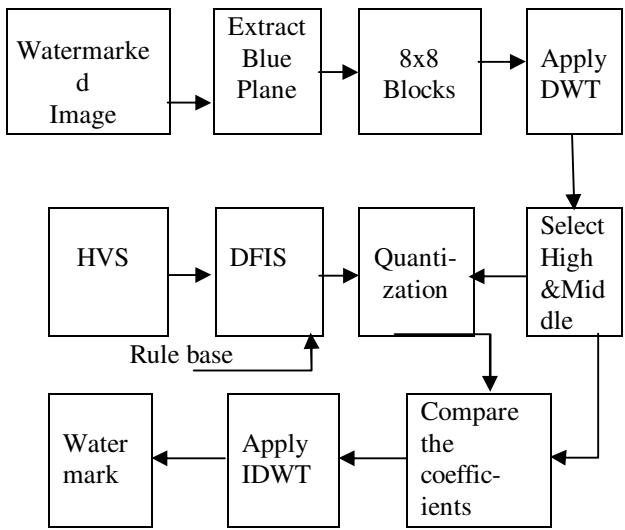

Figure (4): Watermark Extraction

### Watermark Extraction Algorithm:

1.  Select Blue plane of the Watermarked Image.
2.  Divide into 8x8 blocks.
3.  Apply DWT.
4.  Quantize the DWT coefficient T'' (j) by Q and apply to DFIS.
5.  Extract the watermark using the following equation

$$X^t_j = T''(j) - DFIS\left(round\frac{T''(j)}{Q}\right) \quad ... (4)$$

6.  Measure the similarity between the extracted watermarks X'and the original watermark X.

## 6. Experimental Results

The proposed watermarking algorithms are implemented using MATLAB. The imperceptibility and the robustness of the watermarked image are tested with PSNR and NC. Pears image of size512x512 is selected as the cover image. Gray scale bitmap image of size 64x64 Barbara is selected as the watermark. The PSNR of the watermarked image is calculated using the formula

$$PSNR = 10 \log_{10} \frac{(R*R)}{MSE} \quad ......(5)$$

Where R is maximum fluctuation in the cover image=511

$$MSE = \sum_{j=1}^{r} \sum_{k=1}^{c} \frac{[W(j,k) - W^r(j,k)]}{rc} \quad ......(6)$$

Where r = number of rows in the digital image
     c = number of columns in digital image
     w (j,k) = cover image
     w (j,k) = cover image

$$NC = \frac{\sum_j \sum_k [W(j,k)*W^t(j,k)]}{\sum_j \sum_k [W(j,k)*W(j,k)]} \quad ......(7)$$

The performance evaluation of the method is done by measuring imperceptibility and robustness. The normalized correlation coefficient (NC) is used to measure the similarity between the cover image and the watermarked image.  Peak Signal-to-Noise Ratio (PSNR) is used to measure the imperceptibility of the watermarked image. The robustness of the watermarked image is tested by attacks such as JPEG compression, cropping, median filtering, salt & pepper noise attack, and rotation. The robustness of the watermarked image is tested by attacks such as JPEG compression, cropping, median filtering, salt & pepper noise attack, and rotation. The cover image of size 512x512 and watermark of size 64x64 are shown in figure (5). The watermarked image and extracted watermark without any attack are shown in figure (5).

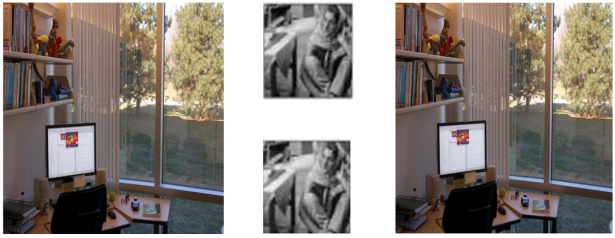

Figure (5): Cover Image, Watermark, Watermarked Image and Extracted Watermark

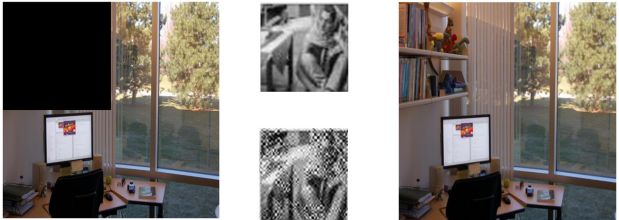

Figure (6): Cropped Image, Extracted Watermark, Median filtered Image and Extracted Watermark.

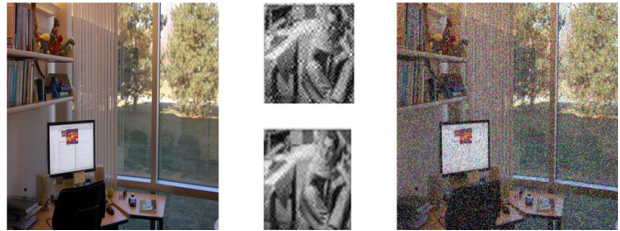

Figure (7): JPEG Attacked Image, Extracted Watermark, Salt&Pepper Noise attacked Image and Extracted Watermark.

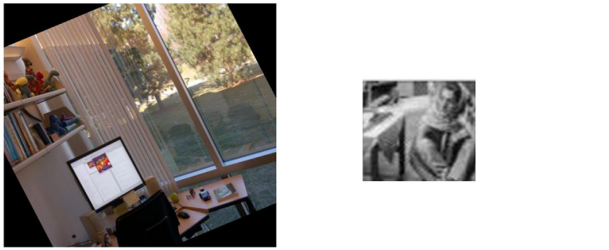

Figure (8): Rotation attacked Image and Extracted Watermark after Rotation attack

| Type of Attack | Intensity | MSE | PSNR | NCC |
|---|---|---|---|---|
| Watermarked Image | -- | 21.1609 | 40.9301 | 0.9965 |
| JPEG Compression | Q=10 | 33 | 38.9818 | 0.9975 |
| Median Filtering | -- | 21 | 40.9272 | 0.6766 |

| Type of Attack | Intensity | MSE | PSNR | NCC |
|---|---|---|---|---|
| Cropping | 5% | 788 | 25.2227 | 0.9736 |
| | 15% | 1289 | 23.0837 | 0.9766 |
| | 25% | 2310 | 20.5498 | 0.9963 |
| | 35% | 3014 | 19.3946 | 0.9960 |
| Salt&Pepper Noise | 5% | 1023 | 24.0884 | 0.9975 |
| | 10% | 2020 | 21.1326 | 0.9975 |
| | 15% | 3057 | 19.3320 | 0.9975 |
| | 20% | 3995 | 18.1700 | 0.9975 |
| Rotation | $4^0$ | 2269 | 20.6275 | 0.9888 |
| | $8^0$ | 3039 | 19.3581 | 0.9877 |
| | $12^0$ | 3499 | 18.7460 | 0.9859 |
| | $16^0$ | 3902 | 18.2720 | 0.9858 |

The cropping attacked image and extracted watermark after cropping are shown in figure (6). The median filtering attacked image and extracted watermark after median filtering are also shown in figure (6). The JPEG compression attacked image and extracted watermark after JPEG compression are shown in figure (7). The salt & pepper noise attacked image and Extracted watermark after salt & pepper noise attack are also shown in Figure (7). The rotation attacked image and extracted watermark after rotation attack are shown in figure (8).

Table 1: Comparison of Various Attacks

The variation of MSE,PSNR,and NCC for different attacks like median filtering, JPEG compression, cropping, salt&pepper noise, and rotation are shown in figures (9),(10), and (11) .

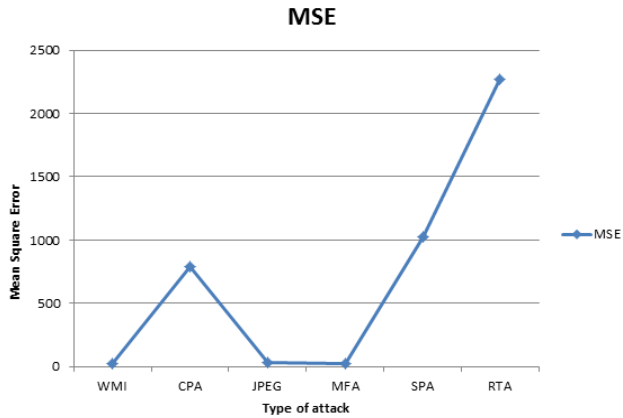

Figure (9): Variation of MSE to different attacks.

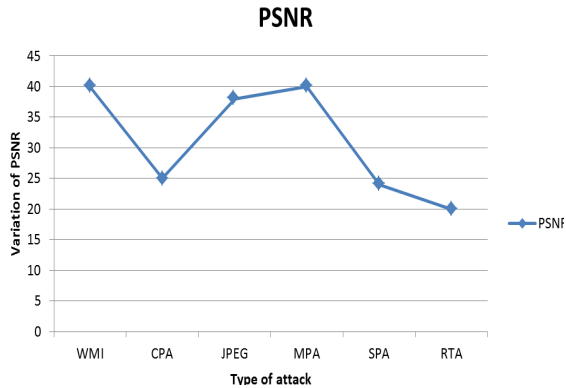

Figure (10): Variation of PSNR to different attacks.

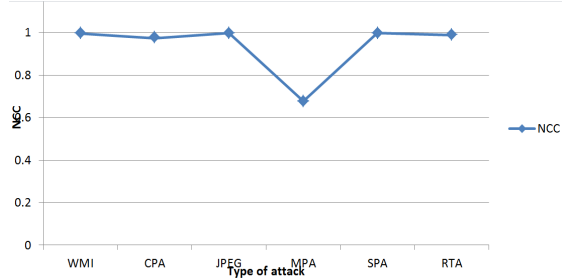

Figure (11): Variation of **NCC** to different attacks.

# 7. Conclusions

In this paper, we proposed the robust digital image watermarking algorithm using quantization and fuzzy logic. The novelty of this algorithm is that the watermark is embedded into high and middle frequency components of the host image using Haar wavelet. That advantage of the proposed mamdani type fuzzy model is that the algorithm is robust to cropping, JPEG compression, salt&pepper noise, and rotation attacks. The drawback is that the algorithm is vulnerable to median filtering attack. The algorithm can also be applied to video images with some modifications.

# References

[1] Nallagarla Ramamurthy and S.Varadrrajan, " Robust Digital Image Watermarking using Quantization and Back propagation Neural Network" ,Contemporary Engineering Sciences, Vol.5,2012,No.3, pp. 137-147.

[2] Nallagarla Ramamurthy and S.Varadrrajan, "Effect of Various Attacks on Watermarked Images", International Journal of Computer Science and Information Technologies, Vol.(3)2, 2012,pp. 3582-3587.

[3] Nizar Sakr, Nicholas.Georganas, and Jiying Zhao, "Copyright Protection of Image Learning Objects using Wavelet based Watermarking and Fuzzy", I2LOR2006,9-11, November,2006.

[4] Samesh Oueslati, et al, "Maximizing Strength of Digital Watermarking using Fuzzylogic", Signal&Image Processing: An International Journal (SIPIJ), Vol.1, No.2, Dec2010, pp.112-124.

[5] Chen Yonqinang, Zhang Yanqing, and Peng Lihua, " A DWT Domain Image Watermarking Scheme Using Genetic Algorithm and Synergetic Neural Network", Academy Publisher,2009, pp. 298-301.

[6] Baisa L.Gunjal and R.RManthalkar, "An overview of transform domain robust digital image watermarking algorithms", Journal of Engineering trends in computing and information sciences, Vol. 2, No. 1, 2010-2011, pp. 37-42.

[7] K. Yogalakshmi and R. Kanchana, "Blind watermarking scheme for digital images "International journal of technology and Engineering systems- Jan-March 2011, Vol 2, No. 3, pp 276-282.

[8] Nagaraj. v ,Dharwadkar, B. B . Amberker, "An Efficient non blind watermarking scheme for colour images using discrete wavelet transformation", International journal of computer applications, Vol. 2, No.3, May 2010, pp. 60-66.

[9] Yanhong Zhang, "Blind Watermark Algorithm Based on HVS and RBF Neural Network in DWT Domain", WSEAS TRANSACTIONS on COMPUTERS, Issue 1, Volume 8, January 2009, pp. 174-183.

[10] Vaishali.S.Jabade, Dr.Sachin R.Gengaje "Literature Review of Wavelet based Digital Image Watermarking Techniques", International Journal of Computer Applications, Vol.31, No.1, October2011.

11.Charu Agarwal, Anurag Mishra, Arpita Sharma " Digital Image Watermarking in DCT Domain using Fuzzy Inference System", IEEE CCECE, pp. 000822-000825.

**Nallagarla Ramamurthy** Received B.Tech in ECE from S.V.University in 1998 and M.Tech in Communication Systems from S.V.University in 2005.Previously worked as an Assistant Professor in Sree Vidyanikethan Engineering College, Tirupati.CurrentlyPursuing Ph.D. from JNTUA, Anantapur. Presented papers in one National Conference and two International Conferences. Published papers in two International journals. Research interests include digital image processing and digital image watermarking.

**Dr.S.Varadrrajan** Received B.Tech degree in ECE from S.V.University in 1987 and M.Tech degree from NIT Warangal, INDIA. He did his Ph.D. in the area of radar signal processing. Currently working as a professor in the Dept. of ECE, S.V.University college of Engineering.